\documentclass{iopart}
\usepackage{euscript,iopams,amssymb,amsfonts,graphicx,bm}
\usepackage{pgfplots,setstack}

\usepackage{float}
\usepackage{epsfig}
\usepackage{esint}

\usepackage{bm,braket}

\eqnobysec

\newcommand{\calU}{{\mathcal U}}

\newcommand{\calP}{{\mathcal P}}
\newcommand{\calQ}{{\mathcal Q}}

\newcommand{\calM}{{\mathcal M}}
\newcommand{\R}{{\mathbb R}}

\newcommand{\X}{\mathbf{X}}
\renewcommand{\P}{\mathbb{P}}
\newcommand{\PP}{\widetilde{P}}
\newcommand{\Q}{\widetilde{Q}}
\newcommand{\calR}{{\mathcal R}}
\newcommand{\J}{\widetilde{J}}

\newcommand{\x}{\mathbf{x}}

\renewcommand{\e}{{\mathrm e}}

\newcommand{\calT}{{\mathcal T}}
\renewcommand{\P}{\mathbb P}
\newcommand{\p}{\widetilde{p}}

\newcommand{\q}{\widetilde{q}}

\renewcommand{\S}{\widetilde{S}}
\newcommand{\ellh}{\widehat{\ell}}

\begin{document}

 \title[Occupation time resetting]{Diffusion in a partially absorbing medium with position and occupation time resetting}

\author{Paul C. Bressloff}
\address{Department of Mathematics, University of Utah 155 South 1400 East, Salt Lake City, UT 84112}

\begin{abstract} 
In this paper we consider diffusion in a domain $\Omega$ containing a partially absorbing target $\calM$ with position and occupation time resetting. The occupation time $A_t$ is a Brownian functional that determines the amount of time that the particle spends in $ \calM$ over the time interval $[0,t]$.
We assume that there exists some internal state $\calU_t$ of the particle at time $t$ which is modified whenever the particle is diffusing within $\calM $. The state $\calU_t$ is taken to be a monotonically increasing function of $A_t$, and absorption occurs as soon as $\calU_t$ crosses some fixed threshold. We first show how to analyze threshold absorption in terms of the joint probability density or generalized propagator $P(\x,a,t|\x_0)$ for the pair $(\X_t,A_t)$ in the case of a non-absorbing substrate $\calM$, where $\X_t$ is the particle position at time $t$ and $\x_0$ is the initial position. 
We then introduce a generalized stochastic resetting protocol in which both the position $\X_t$ and the internal state $\calU_t$ are reset to their initial values, $\X_t\rightarrow \x_0$ and $\calU_t\rightarrow 0$, at a Poisson rate $r$. The latter is mathematically equivalent to resetting the occupation time, $A_t\rightarrow 0$. Since resetting is governed by a renewal process, the survival probability with resetting can be expressed in terms of the survival probability without resetting, which means that the statistics of absorption can be determined by calculating the double Laplace transform of $P(\x,a,t|\x_0)$ with respect to $t$ and $a$. In order to develop the basic theory, we focus on one-dimensional (1D) diffusion with $\calM$ given by a finite or semi-infinite interval, and explore how the MFPT with resetting depends on various model parameters. We also compare the threshold mechanism with the classical case of a constant absorption rate.

\end{abstract}

\maketitle
\section{Introduction}

One of the characteristic features of diffusion in $\R^d$ with stochastic resetting is that the mean first passage time (MFPT) to reach the surface $\partial \calM$ of some target $\calM \subset \R^d$ is rendered finite \cite{Evans11a,Evans11b,Evans14}. Assuming that the position of the particle is reset to its initial location $\x_0\notin \calM$ at a constant rate $r$ (Poissonian resetting), one typically finds that the MFPT is a unimodal function of $r$ with a minimum at an optimal value $r_{\rm opt}$ \cite{Evans11a,Evans11b,Evans14}. The existence of an optimal resetting rate is due to two factors. First, for diffusion in an unbounded domain, the MFPT to reach $\partial \calM$ is infinite in the absence of resetting ($r\rightarrow 0$). This holds irrespective of whether diffusion is recurrent ($d=1,2$) or transient ($d\geq 3$). Second, the MFPT is also infinite in the limit $r\rightarrow \infty$, since resetting to $\x_0$ happens so often that the particle never has enough time to reach $\partial \calM$. The situation changes for diffusion in a bounded domain $\Omega$, since the MFPT is finite in the limit $r\rightarrow 0$. In this case, one observes a phase transition from unimodal to monotonic behavior as the size of the domain $|\Omega|$ is reduced \cite{Pal19}. Analogous results are found for a wide range of stochastic processes with resetting, including non-diffusive processes such as Levy flights \cite{Kus14} and active run and tumble particles \cite{Evans18,Bressloff20}, diffusion in switching environments \cite{Bressloff20a,Bressloff20b,Mercado21} or potential landscapes \cite{Pal15}, resetting followed by a refractory period \cite{Evans19a,Mendez19a}, and resetting with finite return times \cite{Pal19a,Pal19b,Mendez19,Bodrova20,Pal20,Bressloff20c}. (For further generalizations and applications see the review \cite{Evans20} and references therein.) 

Most studies of FPT problems with resetting assume that the target boundary is totally absorbing. Recently we have considered the case of partially absorbing surfaces \cite{Bressloff22c}, in which we adopted a probabilistic model of diffusion-mediated surface reactions developed by Grebenkov \cite{Grebenkov19b,Grebenkov20,Grebenkov22}. The latter exploits the fact that diffusion in a domain with a totally reflecting surface can be implemented probabilistically in terms of so-called reflected Brownian motion, which involves the introduction of a Brownian functional known as the boundary local time \cite{Levy39,McKean75,Majumdar05}. The local time characterizes the amount of time that a Brownian particle spends in the neighborhood of a point on the boundary. In the encounter-based approach, one considers the joint probability density or propagator $P(\x,\ell,t|\x_0)$ for the pair $(\X_t,\ell_t)$ in the case of a perfectly reflecting surface, where $\X_t$ and $\ell_t$ denote the particle position and local time, respectively. The propagator satisfies a corresponding boundary value problem (BVP), which can be derived using integral representations \cite{Grebenkov20} or path integrals \cite{Bressloff22a}. The effects of surface reactions are then incorporated 
 by introducing the stopping time 
${\mathcal T}=\inf\{t>0:\ \ell_t >\widehat{\ell}\}$,
 with $\widehat{\ell}$ a so-called stopping local time \cite{Grebenkov20}. Given the probability distribution $\Psi(\ell) = \P[\ellh>\ell]$, the marginal probability density for particle position is defined according to
 $  p(\x,t|\x_0)=\int_0^{\infty} \Psi(\ell)P(\x,\ell,t|\x_0)d\ell$. It can be shown that the classical Robin boundary condition for the diffusion equation corresponds to the exponential distribution
$\Psi(\ell) =\e^{-\gamma \ell}$, where $\gamma =\kappa_0/D$. (Stochastic resetting for 1D diffusion with a Robin boundary condition has also been analyzed in Ref. \cite{Evans13}.) One natural generalization of the exponential distribution is obtained by taking the reactivity to depend on the local time $\ell$ (or the number of surface encounters), that is, $\kappa=\kappa(\ell)$. A non-trivial feature of partially absorbing surfaces with non-constant reactivities is that one has to maintain a memory of the current boundary local time under reset in order to correctly account for the statistics of absorption. This means that the resetting protocol is not given by a renewal process \cite{Bressloff22b}. In order to recover a renewal process, we introduced a modified resetting rule in which both the position of the particle and the local time were reset. We found that the effects of a partially absorbing surface on the mean first passage time (MFPT) for total absorption increases significantly if local time resetting is included. That is, the MFPT for a totally absorbing surface is increased by a multiplicative factor when the local time is reset, whereas the MFPT is increased additively when only particle position is reset. 

In this paper we extend the theory of diffusion with local time resetting to the case of occupation time resetting. The occupation time $A_t$ is also a Brownian functional, but it determines the amount of time a diffusing particle spends within the interior of a target $\calM$ rather than in a neighborhood of the boundary $\partial \calM$. As we have shown elsewhere \cite{Bressloff22a,Bressloff22b}, the generalized propagator $P(\x,a,t|\x_0)$ for the pair $(\X_t,A_t)$ is the natural analog of the local time propagator in the case of a partially absorbing substrate $\calM$. Here we assume that there exists some internal state $\calU_t$ of the particle at time $t$ which is modified whenever the particle is diffusing within $\calM $. The state $\calU_t$ is taken to be a monotonically increasing function of the occupation time $A_t$, and absorption occurs as soon as $\calU_t$ crosses some fixed threshold. We then introduce a generalized stochastic resetting protocol in which both the position $\X_t$ and the internal state $\calU_t$ are reset to their initial values, $\X_t\rightarrow \x_0$ and $\calU_t\rightarrow 0$, at a Poisson rate $r$. The latter is mathematically equivalent to resetting the occupation time, $A_t\rightarrow 0$. Since resetting is governed by a renewal process, the survival probability with resetting can be expressed in terms of the survival probability without resetting, which means that the statistics of absorption can be determined by calculating the double Laplace transform of $P(\x,a,t|\x_0)$ with respect to $t$ and $a$. (The resetting of an internal state was briefly introduced within the context of local time resetting \cite{Bressloff22c}, but the details of such a mechanism were not explored.) 

 In order to develop the basic theory, we focus on one-dimensional (1D) diffusion and take $\calM$ to be a finite or semi-infinite interval. In section 2, we solve the BVP for the generalized propagator in the absence of resetting, and determine the flux due to absorption within $\calM$. In section 3, we incorporate position and occupation time resetting into the propagator BVP and describe the threshold absorption mechanism. We also use renewal theory to express the survival probability with resetting in terms of the survival probability without resetting. This then allows us to compute the MFPT for absorption with resetting, $T_r$, in terms of the probability flux without resetting, which was calculated in section 2. In section 4 we explore how $T_r$ depends on model parameters, including the resetting rate $r$, the absorption threshold, and the size of the absorbing and non-absorbing regions. We also compare our threshold model with the classical case of a constant absorption rate (the analog of the Robin boundary condition), which was analyzed in some detail elsewhere \cite{Schumm21}. One novel feature of threshold absorption is that the MFPT can exhibit a unimodal (non-monotonic) dependence on $r$ even when the particle starts at the boundary of (or within) the absorbing region.

\section{Partially absorbing interval (no resetting)}

\begin{figure}[t!]
\centering
  \includegraphics[width=10cm]{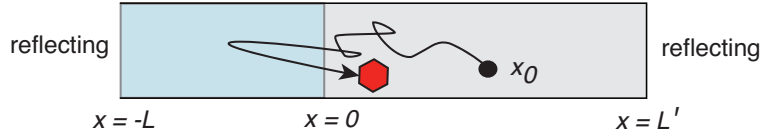}
  \caption{Diffusion in the interval $\Omega=[-L',L]$ with a partially absorbing subinterval $\calM=[-L',0]$.}
  \label{fig1}
\end{figure}

Consider a particle diffusing in the interval $\Omega=[-L,L']$ with a partially absorbing subinterval $\calM=[-L,0]$ and reflecting boundary conditions at $x=-L,L'$, see Fig. \ref{fig1}. Following Refs. \cite{Bressloff22a,Bressloff22b}, we model the absorption process in terms of a generalized propagator, which is the joint probability density for particle position $X_t$ and the so-called occupation time $A_t$ in the absence of absorption. (This is a natural extension of the corresponding propagator for surface-based absorption, which involves the boundary local time \cite{Grebenkov20,Grebenkov22}.) The occupation time is a Brownian functional \cite{Majumdar05} defined according to 
\begin{equation}
\label{occ}
A_t=\int_{0}^tI_{\calM}(X_{\tau})d\tau .
\end{equation}
Here $I_{\calM}(x)$ denotes the indicator function of the set $\calM\subset \Omega$, that is, $I_{\calM}(x)=1$ if $x\in \calM$ and is zero otherwise. Hence, $A_t$ specifies the amount of time the particle spends within $\calM$ over the time interval $[0,t]$. We also take $X_0=x_0$ and $A_0=0$.
Denoting the generalized propagator by $P(x,a,t|x_0)$ and $Q(x,a,t|x_0)$ for $x\in [0,L']$ and $x\in [-L,0]$, respectively, we have the BVP \cite{Bressloff22a}
\numparts
\begin{eqnarray}
\label{Pocca}
\fl &\frac{\partial P(x,a,t|x_0)}{\partial t}=D\frac{\partial^2 P(x,a,t|x_0)}{\partial x^2}, \ 0<x<L',\\
\fl &\frac{\partial Q(x,a,t|x_0)}{\partial t}=D\frac{\partial^2 Q(x,a,t|x_0)}{\partial x^2} ,\nonumber \\ \fl &\quad -\left (\frac{\partial Q}{\partial a}(x,a,t|x_0) +\delta(a)Q(x,0,t|x_0) \right ),\ -L<x<0,
\label{Poccb}\\
\fl &\left .\frac{\partial P(x,a,t|x_0)}{\partial x}\right |_{x=L'}=0,\quad \left .\frac{\partial Q(x,a,t|x_0)}{\partial x}\right |_{x=-L}=0.
\label{Poccc}
\end{eqnarray}
These are supplemented by matching conditions at the interface $x=0$,
	\begin{equation}
	\label{Poccd}
	\fl P(0,a,t|x_0)=Q(0,a,t|x_0),\quad  \left .\frac{\partial P(x,a,t|x_0)}{\partial x}\right |_{x=0}=\left .\frac{\partial Q(x,a,t|x_0)}{\partial x}\right |_{x=0},
	\end{equation}
	\endnumparts
	and the initial conditions $P(x,a,0|x_0)=\delta(x-x_0)\delta(a)$, $Q(x,a,0|x_0)=0$. We assume that the particle starts out in the non-absorbing region. (The analysis is easily modified if $x_0\in [-L,0]$.) Next we introduce the stopping time condition
\begin{equation}
\label{TA}
{\mathcal T}=\inf\{t>0:\ A_t >\widehat{A}\},
\end{equation}
where $\widehat{A}$ is a random variable with probability distribution $\Psi(a)$. Heuristically speaking, ${\mathcal T}$ is a random variable that specifies the time of absorption in $[-L,0)$, which is the event that $A_t$ first crosses a randomly generated threshold $\widehat{A}$. The marginal probability density for particle position $X_t $ is then \cite{Bressloff22a}
\numparts
\begin{eqnarray}
\label{peep}
p(x,t|x_0)&=\int_0^{\infty}\Psi(a) P(x,a,t|x_0)da,\ 0\leq x \leq L',\\
q(x,t|x_0)&=\int_0^{\infty}\Psi(a) Q(x,a,t|x_0)da,\ -L \leq x \leq 0.
\end{eqnarray}
\endnumparts

Given the marginal probability densities, we can define the survival probability $S(x_0,t)$ that the particle hasn't been absorbed up to time $t$, given that it started at $x_0$:
\begin{eqnarray}
\label{Socc}
\fl S(x_0,t)&= \int_0^{L'} p(x,t|x_0)dx+\int_{-L}^0 q(x,t|x_0)dx\nonumber \\
\fl &=\int_0^{\infty}\Psi(a)\left [\int_0^{L'} P(x,a,t|x_0)dx+\int_{-L}^0 Q(x,a,t|x_0)dx\right ]da .
\end{eqnarray}
Differentiating both sides with respect to $t$, and using equations (\ref{Pocca})--(\ref{Poccc}) implies that
\begin{eqnarray}
\fl \frac{\partial S(x_0,t)}{\partial t}&=\int_0^{\infty}\Psi(a)\left [\int_0^{L'}D\frac{\partial^2 P(x,a,t|x_0)}{\partial x^2}dx+\int_{-L}^0 D\frac{\partial^2 P(x,a,t|x_0)}{\partial x^2}dx\right ]da\nonumber \\
\fl &\quad - \int_0^{\infty} \Psi(a)\int_{-L}^0 \left [\frac{\partial Q}{\partial a}(x,a,t|x_0) +\delta(a)Q(x,0,t|x_0)\right ]dx\, da.
\label{Qocc}
\end{eqnarray}
Applying the divergence theorem to the first two integrals on the right-hand side, imposing the Neumann boundary condition on $x=-L,L'$ and flux continuity at $x=0$ shows that these two integrals cancel. Hence, using integration by parts we find that
\begin{eqnarray}
\label{SJ}
\frac{\partial S(x_0,t)}{\partial t}=- \int_0^{\infty} \psi(a) \int_{-L}^0Q(x,a,t|x_0)dx\, da =-J(x_0,t),
\end{eqnarray}
where $\psi(a)=-\Psi'(a)$ and $J(x_0,t)$ is the probability flux due to absorption within $[-L,0]$. 
Finally, Laplace transforming equation (\ref{SJ}) with respect to $t$ and noting that $S(x_0,0)=1$ gives
\begin{equation}
\label{QL}
s\widetilde{S}(x_0,s)-1=- \widetilde{J}(x_0,s)
\end{equation}
with $\widetilde{f}(t)\equiv \int_0^{\infty } f(t)\e^{-st}dt$.
The probability density of the stopping time $\calT$, equation (\ref{TA}), is given by $-\partial S/\partial t$ so that the MFPT (if it exists) is
\begin{eqnarray}
\label{MFPT1}
\fl T(x_0)&=-\int_0^{\infty}t\frac{\partial S(x_0,t)}{\partial t}dt =\int_0^{\infty} S(x_0,t)dt =\S(x_0,0)=-\left .\frac{\partial \widetilde{J}(x_0,s)}{\partial s}\right |_{s=0}.
\end{eqnarray}
Similarly, higher order moments of the FPT density can be obtained in terms of higher order derivatives of $\widetilde{J}(x_0,s)$. 

As we have shown elsewhere \cite{Bressloff22b}, the simplest method for solving the propagator BVP is to 
take a double Laplace transform with respect to both $t$ and $a$ by setting
\numparts
\begin{eqnarray}
\calP(x,z,s|x_0)&=\int_0^{\infty} \e^{-za}\left [\int_0^{\infty} \e^{-st}P(x,a,t|x_0)dt\right ]da,\\ \calQ(x,z,s|x_0)&=\int_0^{\infty} \e^{-za}\left [\int_0^{\infty} \e^{-st}Q(x,a,t|x_0)dt\right ]da,
\end{eqnarray}
\endnumparts
This yields the BVP
\numparts
\begin{eqnarray}
\label{PoccLTa}
\fl & D\frac{\partial^2 \calP(x,z,s|x_0)}{\partial x^2}-s\calP(x,z,s|x_0)=-\delta(x-x_0),\ 0<x<L' , \\
\fl & D\frac{\partial^2 \calQ(x,z,s|x_0)}{\partial x^2}-(s+z)\calQ(x,z,s|x_0)=0,\ -L<x<0 \label{PoccLTb}\\
\fl  \fl &\left .\frac{\partial \calP(x,z,s|x_0)}{\partial x}\right |_{x=L'}=0,\quad \left .\frac{\partial \calQ(x,z,s|x_0)}{\partial x}\right |_{x=-L}=0,
\label{PoccLTc} \\
\fl	&\calP(0,z,s|x_0)=\calQ(0,z,s|x_0),\ \left .\frac{\partial \calP(x,z,s|x_0)}{\partial x}\right |_{x=0}= \left .\frac{\partial \calQ(x,z,s|x_0)}{\partial x}\right |_{x=0}.
\label{PoccLTd}
	\end{eqnarray}
	\endnumparts	
The corresponding solution is \cite{Bressloff22b}
\numparts
\begin{eqnarray}
\label{squida}
\fl  \calP(x,z,s|x_0)&=\frac{1}{\Phi(z,s)D}\frac{\cosh\alpha(s) (L'-x)\cosh\alpha(s) (L'-x_0)}{ \cosh^2\alpha(s) L'}+ G_1(x, s|x_0), \\
\label{squidb}
\fl  \calQ(x,z,s|x_0)&=\frac{1}{\Phi(z,s)D}\frac{\cosh\alpha(s+z) (L+x)}{ \cosh\alpha(s+z) L}\frac{\cosh\alpha(s) (L'-x_0)}{\cosh\alpha(s) L'}, 
\end{eqnarray}
\endnumparts
where $\alpha(s)=\sqrt{s/D}$, 
\begin{equation}
\label{Phi}
\Phi(z,s)\equiv  \alpha(s) \tanh [\alpha(s)L'] +\alpha(s+z) \tanh [\alpha(s+z)L] .
\end{equation}
and $G_1$ is the 1D Green's function that satisfies equation (\ref{PoccLTa}) with a Dirichlet boundary condition at $x = 0$
and a Neumann boundary condition at $x=L'$. In particular,
\begin{eqnarray}
\fl G_1(x, s| x_0) 
    &= \frac{H(x_0 - x)g(x, s)\widehat{g}(x_0, s) +H(x - x_0)g(x_0, s)\widehat{g}(x, s)}{\sqrt{sD}\cosh(\sqrt{s/D}L')},
\end{eqnarray}
where $H(x)$ is the Heaviside function and
\begin{eqnarray}
   g(x, s) = \sinh \sqrt{s/D} x,\quad \makebox{and} \quad \widehat{g}(x, s) =\cosh\sqrt{s/D} (L'-x). 
\end{eqnarray}

Suppose that $z=\kappa_0$ for some fixed rate $\kappa_0$. Equations (\ref{PoccLTa})--(\ref{PoccLTd}) are then equivalent to a classical BVP for diffusion with a constant rate of absorption $\kappa_0$ within $\calM$. In other words, a constant rate of absorption is equivalent to an exponential distribution $\Psi(a)=\e^{-\kappa_0 a}$. (This is analogous to the equivalence of the classical Robin BVP and a partially absorbing surface with an exponential stopping local time distribution \cite{Grebenkov20}.) Hence, we can identify $\calP(x,\kappa_0,t|x_0)$ and $\calQ(x,\kappa_0,t|x_0)$ as the marginal probability densities when the absorption rate is constant, and
\begin{equation}
\label{JLT22}
\widetilde{J}(x_0,s)= \int_{-L}^{0}\calQ(x,\kappa_0,s|x_0) dx .
\end{equation}
 However, various absorption processes are better modeled in terms of a reactivity that is a function of the amount of time a particle spends in contact with a substrate \cite{Bartholomew01,Filoche08}. 
That is, the substrate may be progressively
activated by repeated encounters with a diffusing
particle, or an initially highly reactive surface may become less active due to multiple interactions with the particle (passivation). Both of these cases can be modeled by considering an $a$-dependent reactivity $\kappa=\kappa(a)$ with an associated stopping occupation time distribution
\begin{equation}
\Psi(a)=\exp \left (-\int_0^a\kappa(a')da'\right ).
\end{equation}
Now we have to invert with respect to $z$ to obtain the flux:
\begin{equation}
\label{JLT2}
\widetilde{J}(x_0,s)=\int_0^{\infty} \psi(a)\left [\int_{-L}^{0}{\mathcal L}_{a}^{-1}[\calQ(x,z,s|x_0)] dx \right ]da.
\end{equation}
(Since we will ultimately be interested in FPT moments, there is no need to invert with respect to $s$.)

\section{Position and occupation time resetting}

\begin{figure}[b!]
  \raggedleft
  \includegraphics[width=10cm]{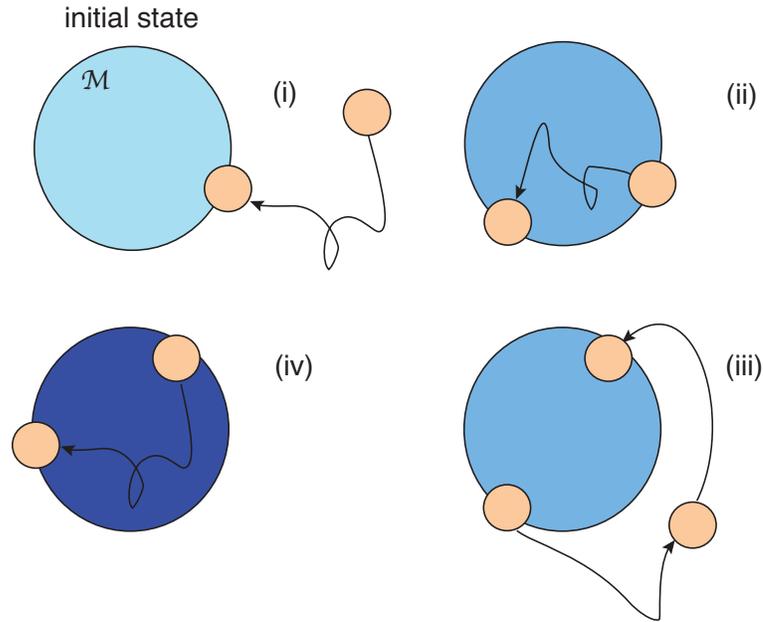}
  \caption{Mechanism for non-constant reactivity within a bounded domain $\calM$, in which each interval of particle-substrate interactions modifies the state of the substrate (as indicated by the substrate shade). (i,iii) Particle diffuses outside $\calM$ so there is no increase in the occupation time nor change of state. (ii,iv) Particle diffuses within $\calM$ such that the occupation time increases and there is a corresponding change of state}
  \label{fig2}
\end{figure}

\begin{figure}[t!]
  \raggedleft
  \includegraphics[width=10cm]{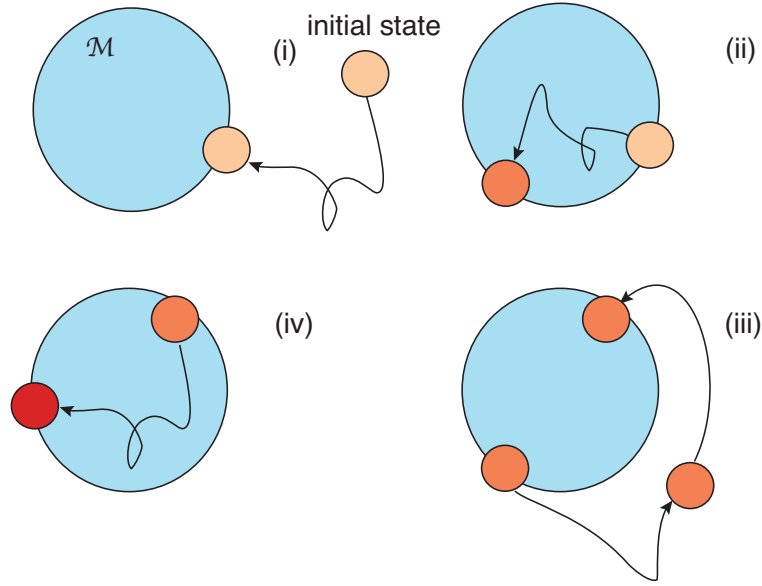}
  \caption{Same as Fig. \ref{fig2} except that each interval of particle-substrate interactions modifies the state of the particle rather than the substrate (as indicated by the particle shade).}
  \label{fig3}
\end{figure}

Following along analogous lines to our recent work on partially reactive boundaries \cite{Bressloff22c}, we modify the propagator BVP (\ref{Pocca})--(\ref{Poccd}) in order to incorporate the effects of stochastic resetting.
Suppose that, prior to absorption, the pair $(X_t,A_t)$ instantaneously resets to the state $(x_0,0)$ at random times determined by a Poisson process with rate $r$, after which diffusion is immediately resumed. The instantaneous resetting of particle position has been studied in a wide variety of stochastic processes \cite{Evans20}, and is an idealized version of a random intermittent search process in which the ballistic or active phase is instantaneous. The interpretation of occupation time resetting (or local time resetting in the case of partially reactive boundaries) is less obvious. As we highlighted at the end of section 2, one of the motivations for considering generalized models of absorption is that the diffusing particle can alter the reactivity of the absorbing substrate. This is illustrated in Fig. \ref{fig2} for a partially absorbing substrate $\calM\subset \R^2$. However, another possible scenario is that the substrate modifies an internal state of the particle, which in turn affects the probability of absorption, see Fig. 3. (One could also imagine a combination of the scenarios shown in Figs. \ref{fig2} and \ref{fig3}.) More specifically, suppose that the particle has some internal state $\calU_t=\calU(A_t)$ with $\calU(0)=0$ and $\calU'(a)> 0$ for all $a\geq 0$. That is, $\calU_t$ is a monotonically increasing function of the occupation time.  Now introduce the stopping time condition
\begin{equation}
\label{TA}
{\mathcal T}=\inf\{t>0:\ \calU_t >\widehat{\calU}\},
\end{equation}
where $\widehat{\calU}$ is a random variable with probability density $p(u)$. The random variable ${\mathcal T}$ specifies the time of absorption within $\calM$, which corresponds to the event that the internal state $\calU_t$ first crosses the randomly generated threshold $\widehat{\calU}$. The marginal probability density for particle position $X_t $ is then
\[p(x,t|x_0)d\x=\P[X_t \in (x,x+dx), \ t < {\mathcal T}|X_0=x_0].\]
If $\calU_t$ is a nondecreasing process, then the condition $t < {\mathcal T}$ is equivalent to the condition $\calU_t <\widehat{\calU}$. This implies that 
\begin{eqnarray*}
p(x,t|x_0)dx&=\P[X_t \in (x,x+dx), \ \calU_t < \widehat{\calU}|X_0=x_0]\\
&=\int_0^{\infty} du\,  p(u)\P[X_t \in (x,x+dx), \ \calU_t < u |X_0=x_0]\\
&=\int_0^{\infty} du\,  p(u)\P[X_t \in (x,x+dx), \ A_t < \calU^{-1}(u) |X_0=x_0]\\
&=\int_0^{\infty} du \ p(u)\int_0^{\calU^{-1}(u)} da' P(x,a',t|x_0)\, dx\\
&=\int_0^{\infty} da \ \frac{p(\calU(a))}{\calU'(a)}\int_0^{a} da' P(x,a',t|x_0)\, dx
\end{eqnarray*}
Finally, using the identity
\[\int_0^{\infty}du\ f(u)\int_0^u du' \ g(u')=\int_0^{\infty}du' \ g(u')\int_{u'}^{\infty} du \ f(u)\]
for arbitrary integrable functions $f,g$, we have
\begin{eqnarray}
\label{peep}
p(x,t|x_0)&=\int_0^{\infty}\Psi(a) P(x,a,t|x_0)da,
\end{eqnarray}
with
\begin{eqnarray}
\psi(a)=-\Psi'(a)=\frac{p(\calU(a))}{\calU'(a)}.
\end{eqnarray}
One could take $\psi(a)$ to be one of the distributions considered in previous studies \cite{Grebenkov20,Bressloff22a}, such as the gamma and Paretto-II distributions. An alternative choice, which we will focus on in this paper, is to assume that absorption takes place as soon as the internal state $\calU_t$ crosses a threshold $u_h$. That is, $p(u)=\delta(u_h-u)$. Setting $a_h=\calU^{-1}(u_h)$, we have $\psi(a)= \delta(\calU(a_h)-u)/\calU'(a)=\delta(a_h-a)$.

\begin{figure}[t!]
  \raggedleft
  \includegraphics[width=12cm]{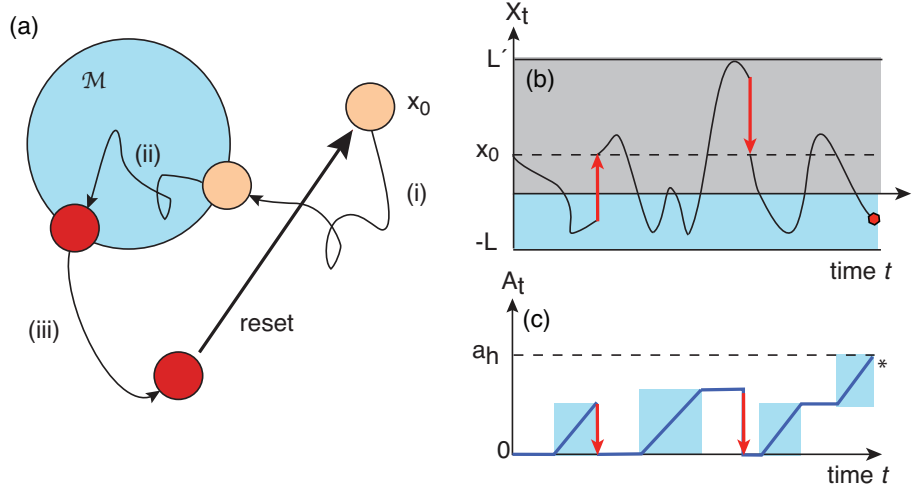}
  \caption{(a) Same as Fig. \ref{fig3} except that now the particle can randomly reset its position and internal state $\calU_t$ at a rate $r$. (b) Schematic illustration of the time course of position $X_t\in [-L,L']$ under the resetting rule $X_t\rightarrow x_0$. (c) Corresponding time course of the occupation time $A_t$ under the effective resetting rule $A_t\rightarrow 0$. The particle is finally absorbed when $A_t$ crosses the threshold $a_h$ (indicated by an asterix).}
  \label{fig4}
\end{figure}

Given the scenario shown in Fig. \ref{fig3}, we can now introduce the aforementioned resetting protocol: whenever the particle resets to $x_0$, its internal state $\calU_t$ is also reset to its initial value, see Fig. \ref{fig4}(a). Mathematically speaking, this is equivalent to resetting the occupation time, $A_t\rightarrow 0$. If this is combined with the threshold absorption mechanism, then a schematic time course of $X_t$ and $A_t$ for the 1D configuration in Fig. \ref{fig1} is of the form shown in Fig. \ref{fig4}(b,c).
Let $P_r(x,a,t|x_0)$ and $Q_r(x,a,t|x_0)$ denote the propagator with resetting outside and inside $\calM$, respectively, in the case of a non-absorbing target. The forward equations with resetting take the form
\numparts
\begin{eqnarray}
\label{rPocca}
\fl &\frac{\partial P_r(x,a,t|x_0)}{\partial t}=D\frac{\partial^2 P_r(x,a,t|x_0)}{\partial x^2}-rP_r(x,a,t|x_0)+r\delta(x-x_0)\delta(a), \ 0<x<L'. \nonumber \\ \fl &\\
\label{rPoccb}
\fl &\frac{\partial Q_r(x,a,t|x_0)}{\partial t}=D\frac{\partial^2 Q_r(x,a,t|x_0)}{\partial x^2} -rQ_r(x,a,t|x_0),\\ \fl &\quad -\left (\frac{\partial Q_r}{\partial a}(x,a,t|x_0) +\delta(a)Q_r(x,0,t|x_0) \right ),\ -L<x<0,\nonumber 
\\
\fl & \left .\frac{\partial P_r(x,a,t|x_0)}{\partial x}\right |_{x=L'}=0,\quad \left .\frac{\partial Q_r(x,a,t|x_0)}{\partial x}\right |_{x=-L}=0
\label{rPoccc}\\
\fl & P_r(0,a,t|x_0)=Q_r(0,a,t|x_0),\quad  \left .\frac{\partial P_r(x,a,t|x_0)}{\partial x}\right |_{x=0}=\left .\frac{\partial Q_r(x,a,t|x_0)}{\partial x}\right |_{x=0}.
\label{rPoccd}
\end{eqnarray}
	\endnumparts	
The additional terms $-rP_r(x,a,t|x_0)$ and $-rQ_r(x,a,t|x_0)$ in equations (\ref{rPocca}) and (\ref{rPoccb}), respectively, represent the loss of probability at each point $x\in [0,L']$ and $x\in [-L,0]$ due to resetting at a uniform rate $r$. This leads to a net flux into the reset point $x_0 \in [0,L]$.
Performing the double Laplace transform along similar lines to the BVP without resetting, we have
\numparts
\begin{eqnarray}
\label{zLTra}
\fl & D\frac{\partial^2 \calP_r(x,z,s|x_0)}{\partial x^2}-(r+s)\calP_r(x,z,s|x_0)=-\left [1+\frac{r}{s}\right ]\delta(x-x_0),\ 0<x<L' , \\
\fl & D\frac{\partial^2 \calQ_r(x,z,s|x_0)}{\partial x^2}-(s+r+z)\calQ_r(x,z,s|x_0)=0,\ -L<x<0\label{zLTrb}\\
\fl  &\left .\frac{\partial \calP_r(x,z,s|x_0)}{\partial x}\right |_{x=L'}=0,\quad \left .\frac{\partial \calQ_r(x,z,s|x_0)}{\partial x}\right |_{x=-L}=0,
\label{zLTrc} \\
\fl	&\calP_r(0,z,s|x_0)=\calQ_r(0,z,s|x_0),\ \left .\frac{\partial \calP_r(x,z,s|x_0)}{\partial x}\right |_{x=0}= \left .\frac{\partial \calQ_r(x,z,s|x_0)}{\partial x}\right |_{x=0}
\label{zLTrd}
	\end{eqnarray}
	\endnumparts

It immediately follows that the resetting protocol is given by a renewal process with respect to the full state space $(X_t,A_t)$. This implies that the propagator satisfies a first renewal equation of the form
\begin{equation}
\label{who}
\fl  Q_r(x,a,t|x_0)= \e^{-rt}Q(x,a,t|x_0)+r\int_0^t\e^{-r \tau}Q_r(x,a,t-\tau|x_0)d\tau,\quad -L<x<0,
\end{equation}
and similarly for $P_r(x,a,t|x_0)$ with $x \in [0,L']$. (The propagators $Q$ and $P$ are the solutions to the BVP without resetting, see equations (\ref{Pocca})--(\ref{Poccd}).)
The first term on the right-hand side of equation (\ref{who}) represents all trajectories that do not undergo any resettings, which occurs with probability $\e^{-rt}$. The second term represents the complementary set of trajectories that reset at least once with the first reset occurring at time $\tau$. Laplace transforming the renewal equation and rearranging shows that
\begin{equation}
\label{con}
\Q_r(x,a,s|x_0)=\left (1+\frac{r}{s}\right )\Q(x,a,r+s|x_0),
\end{equation}
Multiplying both sides of equation (\ref{con}) by $s$ and taking the limit $s\rightarrow 0$, then establishes that there exists a non-equilibrium stationary state (NESS) 
\begin{equation}
\label{con0}
\fl  Q_r^*(x,a|x_0)=\lim_{t\rightarrow \infty}Q_r(x,a,t|x_0)=\lim_{s\rightarrow 0}s \Q_r(x,a,s|x_0)=r\Q(x,a,r|x_0).
\end{equation}
Similarly, $P_r^*(x,a|x_0)=r\PP(x,a,r|x_0)$.

In contrast to the case of partial absorption without resetting, we cannot simply take
$p_r(x,t|x_0)=\int_0^{\infty}\Psi(a)\calP_r(x,a,t|x_0)da$ and $q_r(x,t|x_0)=\int_0^{\infty}\Psi(a)\calQ_r(x,a,t|x_0)da$ since $A_t$ is no longer a monotonically increasing function of time $t$. (A similar issue holds for partially absorbing surfaces and the boundary local time \cite{Bressloff22c}.) Therefore, we first partition the set of contributing paths according to the number of resettings:
\begin{eqnarray}
\fl p_r(x,t|x_0)dx&=\e^{-rt}\P[X_t \in [x,x+dx]|X_0=x_0,\, {\mathcal T}>t,\, {\mathcal I}_t=0]\\
\fl &\quad +r\e^{-rt}\P[X_t \in [x,x+dx]|X_0=x_0,\, {\mathcal T}>t,\, {\mathcal I}_t=1]\nonumber \\
\fl &\quad +r^2\e^{-rt}\P[X_t \in [x,x+dx]|X_0=x_0,\, {\mathcal T}>t,\, {\mathcal I}_t=2]+\ldots \nonumber 
\end{eqnarray}
Here ${\mathcal T}=\inf\{t>0, A_t >\widehat{A}\}$, see equation (\ref{TA}), and ${\mathcal I}_t$ denotes the number of resettings in the interval $[0,t]$.
Then, for a given number of resettings, we decompose a path into time intervals over which $A_t$ is monotonically increasing: 
\begin{eqnarray}
\fl p_r (x,t|x_0)&=\e^{-rt} p_0(x,t|x_0)+r\e^{-rt} \int_0^t p_0(x,\tau|x_0)S(x_0,t-\tau)d\tau\\
\fl &\quad +r\e^{-rt} \int_0^t \int_0^{t-\tau}p_0(x,\tau|x_0)S(x_0,t-\tau)S(x_0,t-\tau-\tau')d\tau'd\tau+\ldots \nonumber
\end{eqnarray}
where $S$ is the survival probability without resetting. Laplace transforming the above equation and using the convolution theorem shows that
\begin{eqnarray}
 \p_r(x,s|x_0)&=\p(x,r+s|x_0)+r\p(x,r+s|x_0)\S(x_0,r+s)\nonumber \\
&\quad +r^2\p(x,r+s|x_0)\S(x_0,r+s)^2+\ldots
\end{eqnarray}
Summing the geometric series and performing an analogous decomposition of $q_r(x,t|x_0)$ yields the results
\numparts
\begin{eqnarray}
\label{prQ}
\p_r (x,s|x_0)&=\frac{\p(x,r+s|x_0)}{1-r\S(x_0,r+s)},\ 0<x < L'\\  \q_r (x,s|x_0)&=\frac{\q(x,r+s|x_0)}{1-r\S(x_0,r+s)},\ -L<x<0.
\end{eqnarray}
\endnumparts
Finally, integrating with respect to $x$ gives
\begin{eqnarray}
 \S_r(x_0,s)&\equiv \int_0^{L'} \p_r (x,s|x_0)dx+\int_{-L}^0 \q_r (x,s|x_0)dx\nonumber \\
&=\int_0^{L'} \frac{\p(x,r+s|x_0)}{1-r\S(x_0,r+s)}dx+\int_{-L}^0 \frac{\q(x,r+s|x_0)}{1-r\S(x_0,r+s)}dx\nonumber \\
&=\frac{\S(x_0,r+s)}{1-r\S(x_0,r+s)}.
\label{SiR}
\end{eqnarray}

\section{Analysis of the MFPT}

Taking the limit $s\rightarrow 0$ in equation (\ref{SiR}) and denoting the MFPT with resetting by $T_r(x_0)$ shows that
\begin{eqnarray}
\label{Tr}
T_r(x_0)=\frac{\S(x_0,r)}{1-r\S(x_0,r)}=\frac{1-\J(x_0,r)}{r\J(x_0,r)},
\end{eqnarray}
with $\S(x_0,r) =[1-\J(x_0,r)]/{r}$
and $\J(x_0,r)$ given by equation (\ref{JLT2}) with $s=r$. In particular, for the threshold density $\psi(a)=\delta(a-a_h)$, we have
\begin{equation}
\label{JLTh}
\widetilde{J}(x_0,r)= \int_{-L}^{0}{\mathcal L}_{a_h}^{-1}[\calQ(x,z,r|x_0)] dx
\end{equation}
with the Laplace transformed propagator $\calQ(x,z,r|x_0)$ given by the solution (\ref{squidb}). For the sake of illustration, we fix the initial position to be on the interface, that is, $x_0=0$ so that (after dropping the dependence on $x_0$)
\begin{eqnarray}
\fl  \calQ(x,z,r)= \frac{1}{\alpha(r) \tanh [\alpha(r)L'] +\alpha(r+z) \tanh [\alpha(r+z)L] }\frac{\cosh\alpha(r+z) (L+x)}{ D\cosh\alpha(r+z) L}.\nonumber \\
\label{solQQ}
\end{eqnarray}
We will explore the dependence of $T_r=T_r(0)$ on the resetting rate $r$, threshold $a_h$ and the lengths $L,L'$ of the absorbing and non-absorbing intervals. The main step is evaluating the inverse Laplace transform with respect to $z$ along analogous lines to Ref. \cite{Bressloff22b}. We first consider the simpler case of an unbounded absorbing substrate.

\subsection{Unbounded partially absorbing substrate ($L\rightarrow \infty$)}

 \begin{figure}[t!]
\centering
  \includegraphics[width=13cm]{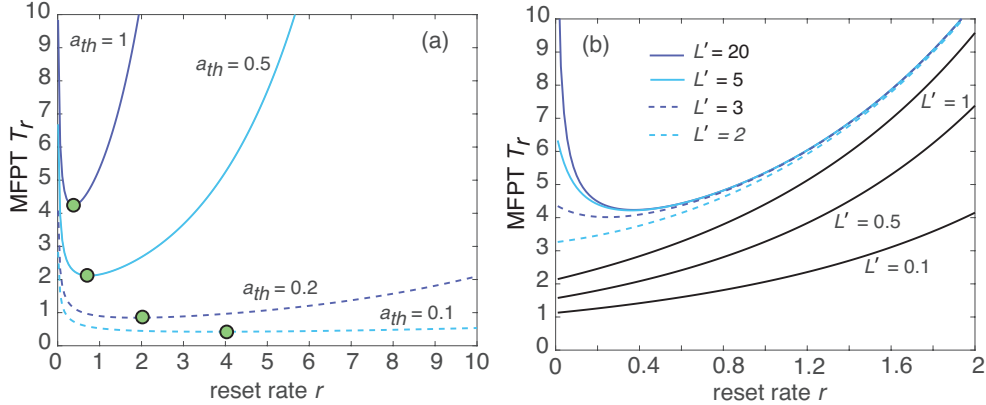}
  \caption{MFPT $T_r$ for an unbounded partially absorbing interval ($L \rightarrow \infty$) and a threshold distribution $\Psi(a)=H(a_{\rm th}-a)$. (a) Plot of $T_r$ as a function of the reset rate for various values of $a_{\rm th}$ and $L'=10$. (b) Corresponding plots for various lengths $L'$ and $a_{\rm th}=1$. We also set $D=1$.}
  \label{fig5}
\end{figure}

In the limit $L\rightarrow \infty$ we have  $\tanh(\sqrt{(s+z)/D}L)\rightarrow 1$ and the $z$-dependence of the solution (\ref{solQQ}) simplifies significantly:
\begin{eqnarray}
\fl  \calQ(x,z,r)= \frac{\e^{\alpha(r+z)x}}{\alpha(r+z)+\alpha(r) \tanh [\alpha(r)L'] }\frac{1}{ D}=\frac{1}{\sqrt{D}}\frac{\e^{-\sqrt{[r+z]/D}|x|}}{\sqrt{r+z}+\sqrt{r}\tanh[\sqrt{r/D}L']}.
\label{solQQ2}
\end{eqnarray}
Integrating with respect to $x$ gives
\begin{eqnarray}
\fl \calR(z,r)\equiv \int_{-\infty}^0\Q(x,z,r)dx= \frac{1}{r+z+\sqrt{r(r+z)}\tanh[\sqrt{r/D}L']}.
\end{eqnarray}
Using a standard table of Laplace transforms, we have the transform pair
\begin{eqnarray}
f(t)=  \e^{k^2t}\mbox{erfc}(k\sqrt{t}),\quad \widetilde{f}(s)=\frac{1}{\sqrt{s}[\sqrt{s}+k]}.
\end{eqnarray}
Hence,
\begin{eqnarray}
 \fl \widetilde{R}(a,r)&= \exp\left (ra\tanh^2[\sqrt{r/D}L']-ra\right )\mbox{erfc}\left (\sqrt{ra}\tanh(\sqrt{r/D}L') \right ).
 \end{eqnarray}
 Combining with equations (\ref{Tr}) and (\ref{JLTh}) then shows that the MFPT for a particle starting at $x_0=0$ is 
 \begin{eqnarray}
\label{Tr2}
T_r =\frac{1-\J(r)}{r\J(r)},\quad \widetilde{J}(r)= \widetilde{R}(a_h,r).
\end{eqnarray}

\begin{figure}[t!]
\raggedleft
  \includegraphics[width=13cm]{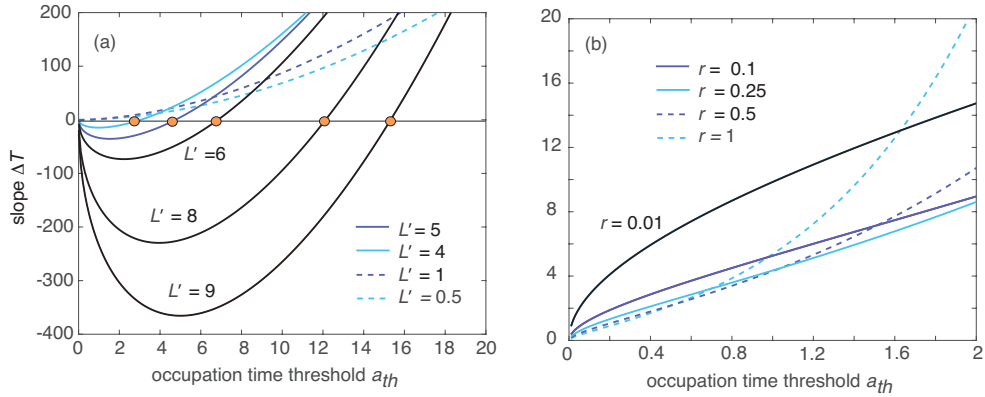}
  \caption{(a) Plot of initial slope $\Delta T=\left .dT_r/dr\right |_{r=0}$ of the MFPT curve as a function of the threshold $a_{\rm th}$ and various length $L'$. The filled circles show the critical threshold beyond which unimodal behavior is lost. (b) Plot of $T_r$ versus $a_{\rm th}$ for various resetting rates and $L'=10$. Other parameters are as in Fig. \ref{fig5}.}
  \label{fig6}
\end{figure}

Example plots of $T_r$ against the reset rate $r$ are shown in Fig. \ref{fig5} for various choices of the length $L'$ of the non-absorbing domain and the occupation time threshold $a_{\rm th}$. For sufficiently large $L'$, the MFPT plot for a given threshold $a_{\rm th}$ exhibits the standard unimodal shape with a minimum MFPT at some optimal reset rate $r_{\rm opt}$. As expected, the MFPT is an increasing function of $a_{\rm th}$ since the probability of absorption is a decreasing function of $a_{\rm th}$. (This is further illustrated in Fig. \ref{fig6}(b)). We find that $r_{\rm opt}$ is also a decreasing function of $a_{\rm th}$. However, for a given $a_{\rm th}$, reducing the length $L'$ leads to a phase transition from a unimodal MFPT curve to a monotonically increasing curve. This is consistent with previous studies of 1D diffusion in a finite interval with absorbing boundaries \cite{Pal19}. The underlying explanation is that, although diffusion in $\R$ is recurrent rather than transient, the MFPT to reach a target point $\bar{x} \in \R$ is infinite in the absence of resetting. This means that $T_r \rightarrow \infty$ as $r\rightarrow 0$. The existence of a minimum follows from the observation that $T_r\rightarrow \infty$ in the limit $r\rightarrow \infty$, since resetting happens so often that the target location $\bar{x}$ is never reached (assuming $\bar{x}\neq x_0$). For finite $L'$, the MFPT $T_0$ is already finite so that taking $r>0$ does not necessarily reduce the MFPT. This is illustrated in Fig. \ref{fig6}(a) where we plot the initial slope $\Delta T=\left . dT/dr \right |_{r=0}$ of the MFPT curve as a a function of $a_{\rm th}$ and various lengths $L'$. A nonzero optimal resetting rate exists provided that $\Delta T <0$. It can be seen that for sufficiently large $L'$, the curve $\Delta T$ versus $a_{\rm th}$ is unimodal such that $\Delta T$ crosses zero at a critical threshold, beyond which $r_{\rm opt}=0$. This critical threshold increases with $L'$. In Fig. \ref{fig6}(b) we plot $T_r$ as a function of $a_{\rm th}$ for various resetting rates. Note that the $T_r$ curves switch from concave down to concave up as $r$ increases.

 \begin{figure}[t!]
\centering
  \includegraphics[width=13cm]{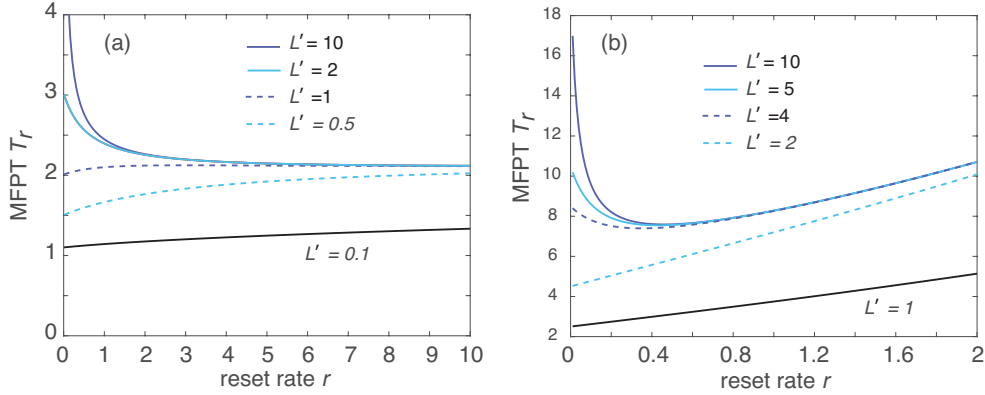}
  \caption{MFPT $T_r$ for an unbounded partially absorbing interval ($L \rightarrow \infty$) and a constant rate of absorption $\kappa_0$ (exponential distribution $\Psi(a)=\e^{-\kappa_0 a/D}$). Plots of $T_r$ as a function of the reset rate for various values of $L'$ and $\kappa_0=1$. (a) $x_0=0$ (b) $x_0=1$. We also set $D=1$.}
  \label{fig7}
\end{figure}

As a point of comparison, in Fig. \ref{fig7} we show MFPT plots for the classical case of a constant absorption rate $\kappa_0$, which is equivalent to the exponential distribution $\Psi(a)=\e^{-\kappa_0 a}$. The MFPT is still given by equation (\ref{Tr}) except that $\widetilde{J}(r)= \kappa_0\calR(\kappa_0,r)$. We find that if $x_0=0$ then there is a major difference in the qualitative behavior compared to the threshold case. That is, in Fig. \ref{fig7}(a) the MFPT curves are monotonic functions of $r$ for all lengths $L'$, switching from monotonically decreasing to monotonically increasing functions as $L'$ is reduced. This is a consequence of the fact that there is no threshold for absorption so that when $L'$ is large, it is beneficial to reset to $x_0$ as fast as possible. This no longer holds when $x_0>0$, where we recover similar behavior to Fig. \ref{fig5}(b), see Fig. \ref{fig7}(b). Finally, note that in the case of threshold absorption, $T_r$ can also exhibit a unimodal dependence on $r$ when the particle starts within the absorbing region $\calM$, that is, $x_0<0$.

\subsection{Bounded partially absorbing substrate ($L<\infty$)}

The $z$-dependence of the solution (\ref{solQQ}) is more complicated when $L$ is finite. For the sake of illustration, let $L'\rightarrow \infty$ so that
(for $x_0=0$)
\begin{eqnarray}
 \calQ(x,z,r)&=\frac{1}{\Phi_{\infty}(z,r)D}\frac{\cosh\alpha(r+z) (L+x)}{ \cosh\alpha(r+z) L},\ -L<x<0,
\end{eqnarray}
where
\begin{equation}
\label{Psi2}
\Phi_{\infty}(z,r)\equiv  \alpha(r)   +\alpha(r+z) \tanh [\alpha(r+z)L] .
\end{equation}
Integrating with respect to $x$ yields
\begin{eqnarray}
 \calR(z,r)&=\frac{1}{\Phi_{\infty}(z,r)D}\frac{\tanh\alpha(r+z) L}{\alpha(r+z) }.
\end{eqnarray}
Following along analogous lines to Ref. \cite{Bressloff22b}, we express the inverse Laplace transform as
\begin{equation}
\label{brom1D}
\widetilde{R}(a,r)=\frac{1}{2\pi i  }\int_{c-i\infty}^{c+i\infty} \e^{za}\frac{1}{\Phi_{\infty}(z,r)D}\frac{\tanh\alpha(r+z) L}{\alpha(r+z) } dz,
\end{equation}
with $c$, $c>0$, chosen so that the Bromwich contour is to the right of all singularities of $\calR(z,r)$. 
Hence, the Bromwich integral (\ref{brom1D}) can be evaluated by closing the contour in the complex $z$-plane and using the Cauchy residue theorem. The resulting contour encloses a countably infinite number of poles, which correspond to the zeros of the function $\Phi_{\infty}(z,s)$. (Since $\alpha(r+z)^{\pm 1}\tanh[\alpha(r+z)L]$ are even functions of $\alpha(r+z)$, it follows that $z=-r$ is not a branch point and $\calR(z,s)$ is single-valued.)

\begin{figure}[t!]
\centering
  \includegraphics[width=13cm]{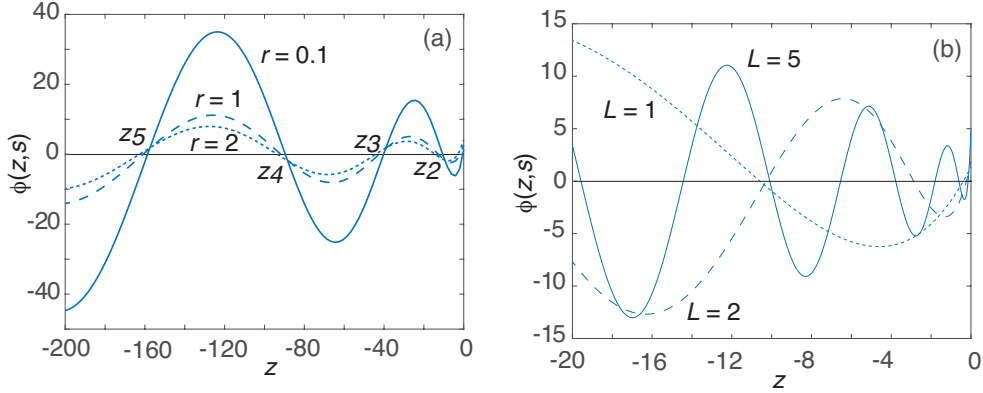}
  \caption{(a) Plot of the function $\phi(z,s)\equiv \Phi_{\infty}(z,r)/\alpha(s)$ against $z$ for various values of $r$ and $L=1$. (b) Corresponding plots for various length $L$ and $r=0.1$. We have also set $D=1$ so that $r_0=1/L^2$.}
  \label{figZ}
\end{figure}

Setting $\Phi_{\infty}(z,r)=0$ in equation (\ref{Psi2}) and rearranging leads to the transcendental equation
\begin{equation}
\label{tran}
\tanh y=-\sqrt{\frac{r}{r_0}}\frac{1}{y},\quad r_0=\frac{D}{L^2},\quad y=L\sqrt{[r+z]/D}.
\end{equation}
Clearly (\ref{tran}) does not have any real solutions. However, there exists a countably infinite number of pure imaginary solutions $y=i\omega_n$, $n\geq 1$,  with $\omega_n$ real such that
\begin{equation}
\label{omm}
\tan \omega_n = \sqrt{\frac{r}{r_0}}\frac{1}{\omega_n}.
\end{equation}
The corresponding zeroes in the $z$-plane are real and lie to the left of $z=-s$ since $\omega_n \neq 0$:
\begin{equation}
z_n=-r-r_0 \omega_n^2,\quad n\geq 1.
\end{equation}
Example plots of the function $\Phi(z,r)/\alpha(r)$ and its zeros are shown in Fig. \ref{figZ}. 
Finally, applying Cauchy's residue theorem, we find that
\begin{equation}
\widetilde{R}(a,r)=\sum_{n\geq 1} \frac{L}{D\partial_z\Phi_{\infty}(z_n,r)}\e^{-(r+r_0\omega_n^2)a}\frac{\tan \omega_n }{\omega_n} ,
\end{equation}
We have used $\alpha(z_n+s)=i\omega_n/L$. Moreover, since $\tan \omega_n/\omega_n= \sqrt{r/r_0}/\omega_n^{2}$ and
\begin{eqnarray}
\partial_z\Phi_{\infty}(z_n,r)&=  \frac{L}{2D }\left (\frac{\tan \omega_n}{\omega_n} +\sec^2 \omega_n \right)\nonumber \\
&=  \frac{L}{2D }\left (1 +\sqrt{\frac{r}{r_0}}\left [1+\sqrt{\frac{r}{r_0}}\right ] \frac{1}{ \omega_n^2} \right),
\label{Psin2}
\end{eqnarray}
it follows that
\begin{equation}
\label{brom1D2}
\widetilde{R}(a,r)=2\sum_{n\geq 1} \frac{\e^{-(r+r_0\omega_n^2)a} }{ \sqrt{\frac{r_0}{r}}\omega_n^2 +\left [1+\sqrt{\frac{r}{r_0}}\right ]  } ,
\end{equation}
which can then be used to determine the MFPT according to equation (\ref{Tr2}).

\begin{figure}[t!]
\centering
  \includegraphics[width=8cm]{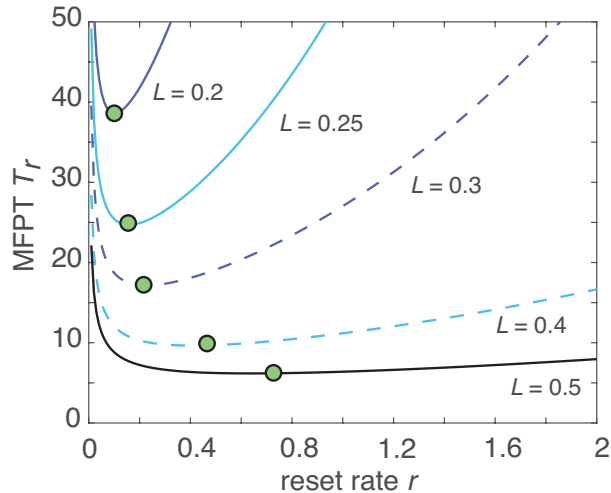}
  \caption{MFPT $T_r$ for a bounded partially absorbing interval ($L< \infty$) and a threshold distribution $\Psi(a)=H(a_{\rm th}-a)$. Plot of $T_r$ as a function of the reset rate for various values of $L$ and $a_{\rm th}=1$. We also set $D=1$ and take $r<r_0$.}
  \label{fig9}
\end{figure}

For the sake of illustration, suppose that $r\ll r_0$ so that the leading order root $\omega_1 \approx 0$. Substituting the approximation $\tan \omega_1\approx \omega_1$ into equation (\ref{omm}) implies that $\omega_1 \approx (r/r_0)^{1/4}$. The higher order roots are approximately equal to the nonzero solutions of $\tan \omega_n=0$, that is, $w_n \approx n \pi$ for $n>1$. In this regime, equation (\ref{brom1D2}) reduces to the more explicit form
\begin{equation}
\label{app1}
\widetilde{R}(a,r)\approx \e^{-\sqrt{r_0r}a} +2\sum_{n\geq 2} \frac{\e^{- r_0n^2\pi^2 a} }{ \sqrt{\frac{r_0}{r}}n^2\pi^2} ,
\end{equation}
In Fig. \ref{fig9} we plot the corresponding MFPT $T_r$ satisfying equation (\ref{Tr}) under the approximation (\ref{app1}). Since the non-absorbing domain is unbounded, we expect $T_r$ to exhibit unimodal behavior and to decrease as the size $L$ of the absorbing interval $\calM$ increases. This is indeed found to be the case.

\section{Conclusion}  

In this paper we considered a hypothetical mechanism for diffusion-mediated absorption, see Figs. \ref{fig3} and Fig. \ref{fig4}, based on the idea that interactions with a target $\calM$ modify an internal state of a diffusing particle, such that once a critical state is reached, the particle is absorbed. This provided the basis of a novel resetting protocol in which both the position and occupation time of the particle are reset. To what extent such an absorption mechanism can be implemented, either naturally or artificially, remains to be seen. However, from a theoretical perspective, it widens the class of resetting protocols to include Brownian functionals. Although we focused on the occupation time $A_t$ associated with the target interior $ \calM$, it is also possible to consider resetting of the corresponding local time $\ell_t$ in cases where the target surface $\partial \calM$ is reactive rather than the interior \cite{Bressloff22c}. As with other models of stochastic resetting \cite{Evans20}, the resetting processes considered here are an idealization of more realistic active processes in which a particle returns to its initial position at some finite speed  \cite{Pal19a,Mendez19,Bodrova20,Pal20}, and there is a refractory period before the particle starts diffusing again \cite{Evans19a,Mendez19a}; the latter could represent the time needed to reset the internal state of the particle. For simplicity, we assumed that resetting is instantaneous and ignored the effects of refractory periods. It would be interesting to explore the effects of delays in future work.

Finally, in this paper we developed the theory of occupation time resetting by considering 1D diffusion, since the solution of the propagator BVP was relatively straightforward. However, in the case of a bounded target $\calM$, it was still necessary to determine the inverse Laplace transform with respect to $z$ by evaluating a Bromwich contour integral. As we have recently shown elsewhere \cite{Bressloff22b}, extending such an analysis to higher-dimensional domains requires computing 
the Laplace transformed propagator in terms of the spectral decomposition of a pair of Dirichlet-to-Neumann operators. The latter reduce to scalars in the 1D case.

\end{document}